\documentclass[11pt,preprint]{aastex}

\usepackage{graphics,graphicx}
\usepackage{natbib}
\usepackage{float}
\usepackage[usenames]{color}

\def\apjl{\rm ApJL}
\def\apjs{\rm ApJS}

\def\aap{\rm AAP}

\newcommand{\gtorder}{\mathrel{\raise.3ex\hbox{$>$}\mkern-14mu
            \lower0.6ex\hbox{$\sim$}}}
\newcommand{\ltorder}{\mathrel{\raise.3ex\hbox{$<$}\mkern-14mu
            \lower0.6ex\hbox{$\sim$}}}
 
\shorttitle{f-modes from highly eccentric double neutron stars}
\shortauthors{Chirenti, Gold and Miller}

\begin{document}

\title{GRAVITATIONAL WAVES FROM F-MODES EXCITED BY THE INSPIRAL OF HIGHLY ECCENTRIC NEUTRON STAR BINARIES}

\author{Cecilia Chirenti\altaffilmark{1}, Roman Gold\altaffilmark{2}, and M. Coleman Miller\altaffilmark{3}}

\affil{
{$^1$}{Centro de Matem\'{a}tica, Computa\c{c}\~{a}o e Cogni\c{c}\~{a}o, UFABC, 09210-170 Santo Andr\'{e}-SP, Brazil}\\
{$^2$}{Perimeter Institute for Theoretical Physics, 31 Caroline Street North, Waterloo, ON N2L 2Y5, Canada}\\
{$^3$}{Department of Astronomy and Joint Space-Science Institute, University of Maryland, College Park, MD 20742-2421 USA}
}

\begin{abstract}

As gravitational wave instrumentation becomes more sensitive, it is interesting to speculate about subtle effects that could be analyzed using upcoming generations of detectors.  One such effect that has great potential for revealing the properties of very dense matter is fluid oscillations of neutron stars.  
These have been found in numerical
simulations of the hypermassive remnants of double neutron star mergers and of highly eccentric neutron star orbits. Here we focus on the latter and 
sketch out some ideas for the production, gravitational-wave detection, and analysis of
neutron star oscillations.
These events will be rare (perhaps up to several tens per year could be detected using third-generation detectors such as the Einstein Telescope or the Cosmic Explorer), but they would have unique diagnostic power for the analysis of cold, catalyzed, dense matter.  Furthermore, these systems are unusual in that analysis of the tidally excited f-modes of the stars could yield simultaneous measurements of their masses, moments of inertia, and tidal Love numbers, using the frequency, damping time, and amplitude of the modes.  They would thus present a nearly unique opportunity to test observationally the I-Love-Q relation.  The analysis of such events will require significant further work in nuclear physics and general relativistic nonlinear mode coupling, and thus we discuss further directions that will need to be pursued.  
For example, we note that for nearly-grazing encounters, numerical simulations show that the energy delivered to the f-modes may be up to two orders of magnitude greater than predicted in the linear theory.

\end{abstract}

\keywords{dense matter --- equation of state --- gravitational waves --- stars: neutron }

\section{INTRODUCTION}
\label{sec:introduction}

The opening of the gravitational wave window to the universe motivates us to dream of what could be accomplished with even more sensitive next-generation detectors.  As an example, proposed third-generation detectors such as the {\it Einstein Telescope} ({\it ET}; \citealt{2010CQGra..27s4002P}) or the {\it Cosmic Explorer} ({\it CE}; \citealt{2016arXiv160708697A}) would have more than an order of magnitude greater sensitivity than Advanced LIGO.  Such an instrument would detect events that would be exceedingly rare with the current generation of detectors, and would therefore give us unique insight into many areas of extreme physics. 

One such area of special interest is the properties of the cold, dense matter in the cores of neutron stars.  This matter has several times the density of atomic nuclei and far more neutrons than protons; that combination cannot be accessed in laboratories, and thus the state of such matter cannot be established by experiments alone.  Nuclear theories diverge widely on the expected composition: possibilities for the dominant component include nucleons, hyperons, or free quarks (see \citealt{2016EPJA...52...66S} for a recent discussion).

Thus constraints on the properties of cold matter beyond nuclear saturation density rely heavily on observations of neutron stars.  Attention has focused in particular on measurements of the gravitational mass $M$ and circumferential radius $R$ of these stars.  Technically, these measurements constrain only the equation of state (e.g., the relation between the pressure $P$ and the total energy density $\epsilon$) rather than the composition, but if $P(\epsilon)$ can be inferred precisely and accurately from $(M,R)$ measurements then that information will be highly valuable to nuclear physicists \citep{1992ApJ...398..569L}.  The discovery of neutron stars with $M\approx 2~M_\odot$ (e.g., PSR~J1614$-$2230 has $M=1.927\pm 0.017~M_\odot$ [\citealt{2016arXiv160300545F}] and PSR~0348+0432 has $M=2.01\pm 0.04~M_\odot$ [\citealt{2013Sci...340..448A}]) has restricted the allowed range of $P(\epsilon)$, but current radius estimates are still dominated by systematic errors \citep{2013arXiv1312.0029M,2016EPJA...52...63M}.

By the early 2030s, when {\it ET} and {\it CE} could take data if they are approved soon \citep{2010CQGra..27s4002P,2016arXiv160708697A}, significantly more information will be available to constrain the properties of cold dense matter.  For example, X-ray measurements from the upcoming {\it NICER} mission (\citealt{2012SPIE.8443E..13G}; the launch is scheduled for 2017) and from the planned {\it LOFT} mission \citep{2012ExA....34..415F,2016RvMP...88b1001W} will constrain neutron star masses and radii. 

Gravitational waveforms from coalescing double neutron star (NS-NS) and neutron star -- black hole (NS-BH) binaries also encode information about the stars.  This is because the static and dynamic tides that are induced in the neutron star(s) take energy from the orbit, which means that the coalescence is more rapid than it would be between two point masses.  The waveforms can therefore be analyzed to determine the tidal deformability and thus aspects of the structure of the stars  \citep{2008PhRvD..77b1502F,2009PhRvD..79l4033R,2010PhRvD..81l3016H,2010PhRvL.105z1101B,2012PhRvD..86d4030B,2013PhRvD..87d4001H,2013PhRvD..88d4042R,2013PhRvL.111g1101D,2014PhRvD..89d3009L,2014PhRvD..89j3012W,2015PhRvL.114p1103B,2015ApJ...804..114B,2015PhRvD..92b3012A,2016PhRvL.116r1101H,2016arXiv160801907S}.  There are important technical issues to overcome, e.g., point-mass waveforms need to be determined to the 4~PN and 4.5~PN orders to avoid biasing tidal extractions \citep{2014PhRvL.112j1101F,2014PhRvD..89b1303Y}, but we assume these will be resolved within the next decade. If a NS-NS merger produces a hypermassive neutron star (which is one that only avoids collapse because of its differential rotation), then several studies have shown that the frequency of the fundamental, or f-mode, pressure oscillation is strongly correlated with the radius of a star whose mass is related to the chirp mass of the binary \citep{2012PhRvL.108a1101B,2012PhRvD..86f3001B,2014PhRvL.113i1104T,2015PhRvD..91f4001T,2014PhRvD..90b3002B,2016EPJA...52...56B}.

Thus by the time that {\it ET} and {\it CE} take data, our understanding of neutron star matter will be improved dramatically from its current state.  However, in terms of direct measurements, only certain combinations of parameters will be available.  X-ray measurements will yield the mass and radius of a few stars to $\sim 5$\% precision \citep{2012SPIE.8443E..13G}.  Observations of the coalescence of neutron stars will give information about the masses of neutron stars and their tidal deformability, quantified by the Love number $\lambda$ \citep{2008PhRvD..77b1502F}.  Detection of f-modes from hypermassive neutron stars could tell us more about the stars, but these will be detectable in only $\sim 0.1$\% of mergers \citep{2012PhRvL.108a1101B} and the analysis of these modes will be highly non-trivial because for such stars differential rotation, dynamically important magnetic fields, and nonzero temperatures could play important roles.  The I-Love-Q relations (\citealt{2013Sci...341..365Y} and subsequent papers) could be used to relate a single observable, such as $\lambda$, to other quantities, but to test the relations observationally it will be necessary to measure at least two of $\lambda$, the moment of inertia $I$, and the rotationally-induced quadrupole $Q$ for individual neutron stars.  It has even been proposed that general relativity itself could be tested if at least two of $I$, $Q$, and $\lambda$ could be measured \citep{2013PhRvD..88b3009Y}.

We therefore consider another scenario, which will have a small rate per volume but which could be seen at a reasonable rate using {\it ET} and {\it CE}.  In this scenario, two neutron stars or a neutron star and a black hole approach each other on highly eccentric orbits, such that at pericenter their separation is of order their radii but not close enough to produce a collision.  In such a situation, oscillations will be induced in the star(s).  Unlike in post-merger hypermassive NS oscillations, the oscillations caused by a close eccentric passage will take place in cold, slowly rotating neutron stars, which means that the modes of the stars will be easier to predict and analyze than they would be in the differentially rotating case.  Thus such oscillations would offer the opportunity to analyze clean systems, and in the same way that asteroseismology has revealed a host of properties of stars, many of the properties of neutron stars could be studied with precision should these oscillations be observed. 

For example, the f-mode frequency $\omega$ and damping time $\tau$ of a compact star are related to $M$ and $I$ by universal (equation of state independent) relations using the ``effective compactness" $\eta = \sqrt{M/I^3}$. These relations were proposed by \cite{2010ApJ...714.1234L},  improving on previous work done by \citealt{Tsui:2005zf} (see also \citealt{Chirenti:2015dda}). Once $\omega$ and $\tau$ are obtained from the gravitational wave observations, the inverse problem can be solved to provide $M$ and $I$.  Given the sensitivity of {\it ET} and {\it CE} we expect that the masses of the binary components will be known from the analysis of the inspiral gravitational wave signal, which means that this independent determination of $M$ will serve as a consistency test.

The amplitude of the tidally excited f-mode will also provide important information, because it will depend on the orbital energy deposited in the mode. This can in turn be related to $\lambda$ \citep{1996ApJ...463..284Q,2008ApJ...677.1216H}.  This means that it is possible to obtain $\lambda$ in the type of system we are considering, and therefore the f-modes detected using {\it ET} and {\it CE} from eccentric neutron star encounters will provide a nearly unique opportunity to measure two of $I$, $\lambda$, and $Q$ (we note that \citealt{2015MNRAS.454.4066P} and \citealt{2016ApJ...818L..11T} show that geodesic and fluid modes of accretion disks around weakly magnetic neutron stars could be affected by both $I$ and $Q$, but the interpretation of the associated quasi-periodic brightness oscillations is far from settled and might not be resolved by the early 2030s).

In Section~\ref{sec:methods} we discuss the expected rates of the required close encounters as well as the oscillations and potential complicating factors.  We discuss our results and present our conclusions in Section~\ref{sec:summary}.

\section{METHODS AND RESULTS}
\label{sec:methods}

\subsection{Estimated Rates}

Because the rates of highly eccentric NS-NS and NS-BH encounters will be low, we will consider only prospective detections using third-generation gravitational wave detectors; for example, {\it ET} would be able to detect NS-NS coalescences out to a redshift $z\sim 2-3$ \citep{2011PhRvD..83b3005Z}, and {\it CE} would be able to detect double neutron star mergers out to $z\sim 6$ \citep{2016arXiv160708697A}.

Because redshifts cannot be neglected for such future detectors, we can use the detection-weighted comoving volume
\begin{equation}
{\bar V}_c\equiv\int_0^\infty{dV_c\over{dz}}f_d(z){1\over{1+z}}dz\; ,
\end{equation}
where $dV_c/dz$ is the differential comoving volume at $z$ and $f_d(z)$ is the probability of detection for an event at $z$ after averaging over sky directions and binary orientation.  The factor $1/(1+z)$ accounts for the difference between observer and source clocks.  Using the ET-D noise spectrum, and assuming that the rest-frame rate per comoving volume is independent of $z$, ${\bar V}_c\approx 28$~Gpc$^{-3}$ for a merger between two $1.4~M_\odot$ neutron stars, which is $\approx 1000$ times the detection volume expected for Advanced LIGO at its design sensitivity (e.g., \citealt{2010CQGra..27q3001A}).  We are grateful to Sebastian Gaebel and Ilya Mandel for suggesting this definition of the volume and for performing the ET-D calculation.  The greater sensitivity of {\it CE} means that its detection-weighted comoving volume would be several times larger still.

No gravitational waves from NS mergers have yet been detected \citep{2016arXiv160707456T}, which means that their rates are highly uncertain (upper limits from the Advanced LIGO O1 run are $\approx 12,600$~Gpc$^{-3}$~yr$^{-1}$ for NS-NS coalescences and $\approx 3,600$~Gpc$^{-3}$~yr$^{-1}$ for NS-BH mergers; see \citealt{2016arXiv160707456T}), and the rates of very eccentric mergers are still more uncertain.  One channel that has been proposed involves two initially mutually unbound neutron stars in a dense stellar system such as a nuclear star cluster, which capture each other via the emission of gravitational radiation during a close passage \citep{2006ApJ...648..411K,2009MNRAS.395.2127O,2012PhRvD..85l3005K,2013ApJ...777..103T}.  \citet{2013ApJ...777..103T} estimates that the volume rate of such encounters is $\sim 0.003-6~{\rm Gpc}^{-3}~{\rm yr}^{-1}$, depending on the degree of segregation of black holes (his equations A20 and A22).  This would suggest that within the horizon of {\it ET} there are $\sim 0.1-200$ mergers per year in this channel, and again {\it CE} would see several times this number.

We are interested in encounters in which the two neutron stars pass within of order their radius from each other at pericenter, because only in such close passes will tidal modes be excited strongly enough to compete with point-mass gravitational radiation as a way of removing energy from the orbit (see the next section).  All hyperbolic encounters of this type are strongly gravitationally focused, which means that the cross section of an orbit scales linearly with the pericenter distance $r_p$.  From \citet{1989ApJ...343..725Q}, the maximum pericenter for gravitational wave capture of two point masses $m_1$ and $m_2$ and thus total mass $m_{\rm tot}=m_1+m_2$ and symmetric mass ratio $\eta\equiv m_1m_2/m_{\rm tot}^2$, which have an initial relative speed at infinity of $v_{\rm rel}$, is
\begin{equation}
r_{p,{\rm max}}=190~{\rm km}\left(\eta\over{0.25}\right)^{2/7}\left(m_{\rm tot}\over{2.8~M_\odot}\right)\left(v_{\rm rel}\over{1000~{\rm km~s}^{-1}}\right)^{-4/7}\; .
\end{equation}
For typical nuclear star cluster velocity dispersions of a few hundred km~s$^{-1}$, $r_{p,{\rm max}}$ is therefore a few hundred kilometers, which means that $\sim (10-15)~{\rm km}/200~{\rm km}\sim{\rm few}$ percent of encounters will have $r_p$ within $\sim 2-3$ neutron star radii, assuming a neutron star radius $R_{\rm NS}\sim 10-15$~km.  Thus from this channel alone we expect ${\rm few}\times (10^{-3}-1)$ NS-NS coalescences per year within the {\it ET} volume, and several times that within the {\it CE} volume, that have pericenter distances of order the neutron star radii.  We also note that for very close encounters, the pericenter distance can decrease significantly from one orbit to the next (see for example the expressions in \citealt{2005PhRvD..72h4009G}), which enhances the rate of encounters that are very close while also being highly eccentric.

Additional enhancements to this number are possible.  For example, Samsing et al. (\citealt{2014ApJ...784...71S}; see also \citealt{2004ApJ...616..221G,2006ApJ...640..156G}) find that the primary channel of eccentric encounters may well involve binary-single interactions, which during the common quasi-resonant phase have multiple opportunities to have very close encounters between neutron stars.  Their work focused on binary-single encounters in globular clusters, but if we extrapolate their work to the higher velocity dispersion environments of nuclear star clusters (which have a smaller fraction of binaries that are on average tighter than those in globular clusters) we could expect an increase by a factor of several compared with the \citet{2013ApJ...777..103T} rate.  There are some difficulties in detecting eccentric binaries relative to circular binaries \citep{2011ApJ...737L...5S,2012PhRvD..85l3005K,2012PhRvD..85l4009E,2013PhRvD..87l7501H,2014PhRvD..90j3001T}, but we perhaps optimistically assume that eccentric templates will be available by the time that {\it ET} is operating, and thus that the extra complexity of eccentric orbits will not reduce the detection rate dramatically.  We note in particular that with pericenter distances as small as we require, the apocenter after first passage will be inside the frequency band of ground-based detectors (e.g., for $r_p\sim 20$~km the apocenter distance will be some tens of gravitational radii), which means that coherent analysis will be possible rather than relying on incoherent addition of gravitational waves from distinct bursts.

Thus the total rate of detection with {\it ET} or {\it CE}, although highly uncertain, seems likely to lie in the range of $\sim 0.1$ per year to several tens per year.  We are interested in the detection of high-frequency oscillation modes rather than the full coalescence, but we note that if a highly eccentric coalescence is discovered then the direction and time of the event will be known precisely, as will the masses of the stars, and that the possible frequency range for f-modes will be relatively small.  Thus the search for the modes will not require nearly the signal to noise ratio that would be needed if the search had no additional information.  We also note that, depending on the exact formation mechanism, it could be that the much greater star formation rate at $z\sim 1-2$ compared with now will enhance the rate per volume of eccentric mergers at that time.  Fundamentally, although it will require {\it ET} or {\it CE} rather than Advanced LIGO and other second-generation detectors to see such events at a reasonable rate, these inspirals will plausibly be detected using third-generation detectors within a few years of operation.  In addition, the lowered observer-frame frequencies for events at $z\sim 1-2$ will also place the high-frequency oscillations at higher-sensitivity frequencies; for example, the expected sensitivity for both {\it ET} and {\it Cosmic Explorer} at $\sim 500$~Hz will be roughly $2-3$ times greater than the sensitivity at the rest-frame $\sim 1500$~Hz frequencies of f-modes (see Figure~1 of \citealt{2016arXiv160708697A}).

\subsection{Excitation of Modes}

Given their typical frequencies of 1.5 kHz, the resonant excitation of f-modes in NS-NS binaries requires either fast rotation of the binary components \citep{1999MNRAS.308..153H} or highly eccentric orbits \citep{1977ApJ...216..914T}. 
When considering highly eccentric binaries, it is useful to work in the limiting case of parabolic orbits with $e = 1$, because that simplifies the calculations. We use the formalism of \citet{1977ApJ...213..183P} to estimate the energy that goes into each of the oscillation modes. The $l = m = 2$ modes of oscillation will emit gravitational waves, and will thus change the orbital dynamics of the system and the total emitted gravitational wave signal of the binary.  Note that this formalism does include the effect of resonances, but that the excitation power spectrum from the tidal interactions is smooth and thus there is no special pericenter distance at which there is an especially strong interaction.  There are two reasons for this: (1)~the Fourier power spectrum of the tidal excitation consists of peaks spaced in frequency by the inverse of the period of the orbit, and this spacing tends to zero in the limit $e\rightarrow 1$, and (2)~changes in the orbit due to gravitational radiation near the pericenter broaden those peaks.  The net result is that the excitation amplitude increases smoothly with decreasing pericenter.

We also note that the \citet{1977ApJ...213..183P} formalism assumes Newtonian orbits, which therefore do not undergo pericenter precession.  This will not be a good approximation for the nearly grazing encounters between neutron stars, or a neutron star and a black hole, which will most strongly excite f-modes.  The strong precession that is expected will increase the time per orbit in which the stars are close to each other, and will therefore increase the tidal coupling to f-modes; this is a situation in which a genuine resonance might play an important role.  Further analysis will be needed to determine the strength of this effect.  Here we simply follow the standard Newtonian treatment.

For a mode with mode numbers $nlm$, the energy transfer to star 1 due to tides from star 2 can be estimated as \citep{1977ApJ...213..183P}:

\begin{equation}
\Delta E_{nlm}=2\pi^2{GM_1^2\over{R_1}}\left(M_2\over{M_1}\right)^2\left(R_1\over{R_{\rm min}}\right)^{2l+2}|Q_{nl}|^2|K_{nlm}|^2\; .
\label{eq:energy}
\end{equation}
Here $M_1$ and $R_1$ are the mass and radius of star 1, $M_2$ is the mass of star 2, and $R_{\rm min}$ is the pericenter distance of the encounter.  $K_{nlm}$ is defined in eq. (\ref{eq:K_nlm}) below, and 
\begin{equation}
Q_{nl}\equiv {1\over{M_1R_1^2}}\int d^3x \rho {\bf\xi}_{nlm}^*\cdot\nabla\left[r^l Y_{lm}(\theta,\phi)\right]\ 
\end{equation}
is the overlap integral between the oscillation mode and the tidal deformation of the star. 

In order to make further analytical progress, we will make the illustrative assumption that the unperturbed star has constant density but that the star is compressible.  We expect that the tidal coupling for real stars will differ quantitatively but not qualitatively from our results.  For such a star (see for example \citealt{1994ApJ...432..296R}), to within a normalizing constant, the Lagrangian displacement field of the order-$l$ f-mode is
\begin{equation}
{\bf\xi}_{nlm}=\nabla\left[r^l Y_{lm}(\theta,\phi)\right]\; ,
\end{equation}
(such that the eigenfunctions are orthogonal with weight $\rho$, the total density of the star) which allows us to write
\begin{equation}
Q_{nl} =  {1\over{M_1R_1^2}}\int d^3x \rho |{\bf\xi}_{nlm}|^2 =  {1\over{M_1R_1^2}}\int d^3x \rho \left|\nabla\left[r^2 Y_{22}(\theta,\phi)\right]\right|^ 2\ 
\label{eq:Q_nl}
\end{equation}
for $l=m=2$. We normalize the spherical harmonics so that
\begin{equation}
Y_{22}={1\over 4}\sqrt{15\over{2\pi}}\sin^2\theta e^{2i\phi}\; .
\end{equation}

Using the gradient operator in spherical coordinates
\begin{equation}
\nabla\psi={\hat e}_r{\partial\psi\over{\partial r}}+{\hat e}_\theta {1\over r}{\partial\psi\over{\partial\theta}}+{\hat e}_\phi{1\over{r\sin\theta}}{\partial\psi\over{\partial\phi}}\; ,
\end{equation}
where the ${\hat e}_i$ are unit vectors along the coordinate direction, we find that
\begin{equation}
\nabla(r^2Y_{22})=\left[2rY_{22}{\hat e}_r+2r\cot\theta Y_{22}{\hat e}_\theta+{2ir\over{\sin\theta}}Y_{22}{\hat e}_\phi\right]\; ,
\end{equation}
and the complex square is 
\begin{equation}
\begin{array}{rl}
\left|\nabla\left[r^2 Y_{22}(\theta,\phi)\right]\right|^ 2&=4r^2|Y_{22}|^2+4r^2\cot^2\theta|Y_{22}|^2+4\csc^2\theta|Y_{22}|^2\\
&={15\over{4\pi}}r^2\sin^2\theta\; .\\
\label{eq:c_square}
\end{array}
\end{equation}

Now we can use eq. (\ref{eq:c_square}) and assume that $\rho=\rho_0$ is a constant to compute the overlap integral (\ref{eq:Q_nl}). After the integration we find
\begin{equation}
Q_{nl} = [(4\pi/3)\rho_0 R_1^3/M_1][3/(2\pi)]=3/(2\pi)\; ,\\
\end{equation}
in good agreement with the overlap integrals calculated by \cite{1995MNRAS.275..301K}. The other factor in the expression for the energy transfer $\Delta E_{nlm}$ (\ref{eq:energy}) is 
\begin{equation}
\label{eq:K_nlm} 
K_{nlm}={W_{lm}\over{2\pi}}2^{3/2}{\hat\eta}I_{lm}({\hat\omega}_{nlm})\; ,
\end{equation}
where 
\begin{equation}
W_{lm}=(-1)^{(l+m)/2}\left[{4\pi\over{2l+1}}(l-m)!(l+m)!\right]^{1/2}\left[2^l\left(l-m\over 2\right)!\left(l+m\over 2\right)!\right]^{-1}=\left(3\pi/10\right)^{1/2}\; ,
\end{equation}
and the last equality holds for $l=m=2$.  Thus $W_{22}$ is very close to unity.
Also in the expression for $K_{nlm}$, ${\hat\eta}\equiv\left[M_1/(M_1+M_2)\right]^{1/2}(R_{\rm min}/R_1)^{3/2}$ and ${\hat\omega}_{nlm}\equiv \omega_{nlm}[G(M_1+M_2)]^{-1/2}R_{\rm min}^{3/2}$ is the mode frequency scaled by the Keplerian frequency of the circular orbit at the pericenter.  Let us specialize to an equal-mass NS-NS binary, $M_1=M_2$, which means that ${\hat\eta}=2^{-1/2}(R_{\rm min}/R_1)^{3/2}$ and thus that
\begin{equation}
K_{nlm}\approx (1/\pi)(R_{\rm min}/R_1)^{3/2}I_{lm}({\hat\omega}_{nlm})\; .
\end{equation}
The integral is 
\begin{equation}
I_{lm}({\hat\omega}_{nlm})=\int_0^\infty(1+x^2)^{-l}\cos\left[2^{1/2}{\hat\omega}_{nlm}(x+x^3/3)+2m\tan^{-1}x\right]dx\; .
\label{eq:Iinteg}
\end{equation}
and the mode frequency for $l=m=2$ of a constant-density star is given by equation 8.102 of \citet{2010ASSL..371.....S}:
\begin{equation}
\omega_{nlm}=\sqrt{{4\pi G\rho\over 3} {2l(l-1)\over{2l+1}}}=\sqrt{(4/5)GM_1/R_1^3} \equiv \omega \label{eq:omega}
\end{equation}
for $n=0$, $l = m = 2$ and thus for an equal-mass binary, 
\begin{equation}
{\hat\omega}_{022}=\sqrt{(2/5)(R_{\rm min}/R_1)^3}\; . \end{equation}
The integral therefore depends on the ratio $R_{\rm min}/R_1$.  Returning to the energy transfer equation (\ref{eq:energy}), we now collect the various factors to find
\begin{equation}
\Delta E_{22}\approx2\pi^2{GM_1^2\over{R_1}}\left(M_2\over{M_1}\right)^2\left(R_1\over{R_{\rm min}}\right)^6|Q_{nl}|^2\left({1\over{\pi^2}}\right)\left({R_{\rm min}\over{R_1}}\right)^3I^2_{22}\; .
\end{equation}
If we say that $M_1=M_2$ and that $Q_{nl}=3/(2\pi)$, and multiply by 2 to take into account that the modes of both stars will be excited, we get
\begin{equation}
\Delta E_{\rm tide,22}={9\over{\pi^2}}{GM_1^2\over{R_1}}\left(R_1\over{R_{\rm min}}\right)^3I^2_{22}\; .
\end{equation}

We want to compare this with the energy radiated in one orbit due to gravitational radiation from the two point masses.  From \citet{1964PhRv..136.1224P} the orbit-averaged energy loss rate is
\begin{equation}
\biggl\langle{dE\over{dt}}\biggr\rangle=-{32\over 5}{G^4M_1^2M_2^2(M_1+M_2)\over{c^5a^5(1-e^2)^{7/2}}}\left(1+{73\over{24}}e^2+{37\over{96}}e^4\right) 
\end{equation}
for a binary of semimajor axis $a$ and eccentricity $e$.

\begin{figure}[!htb]
\begin{center}
\plotone{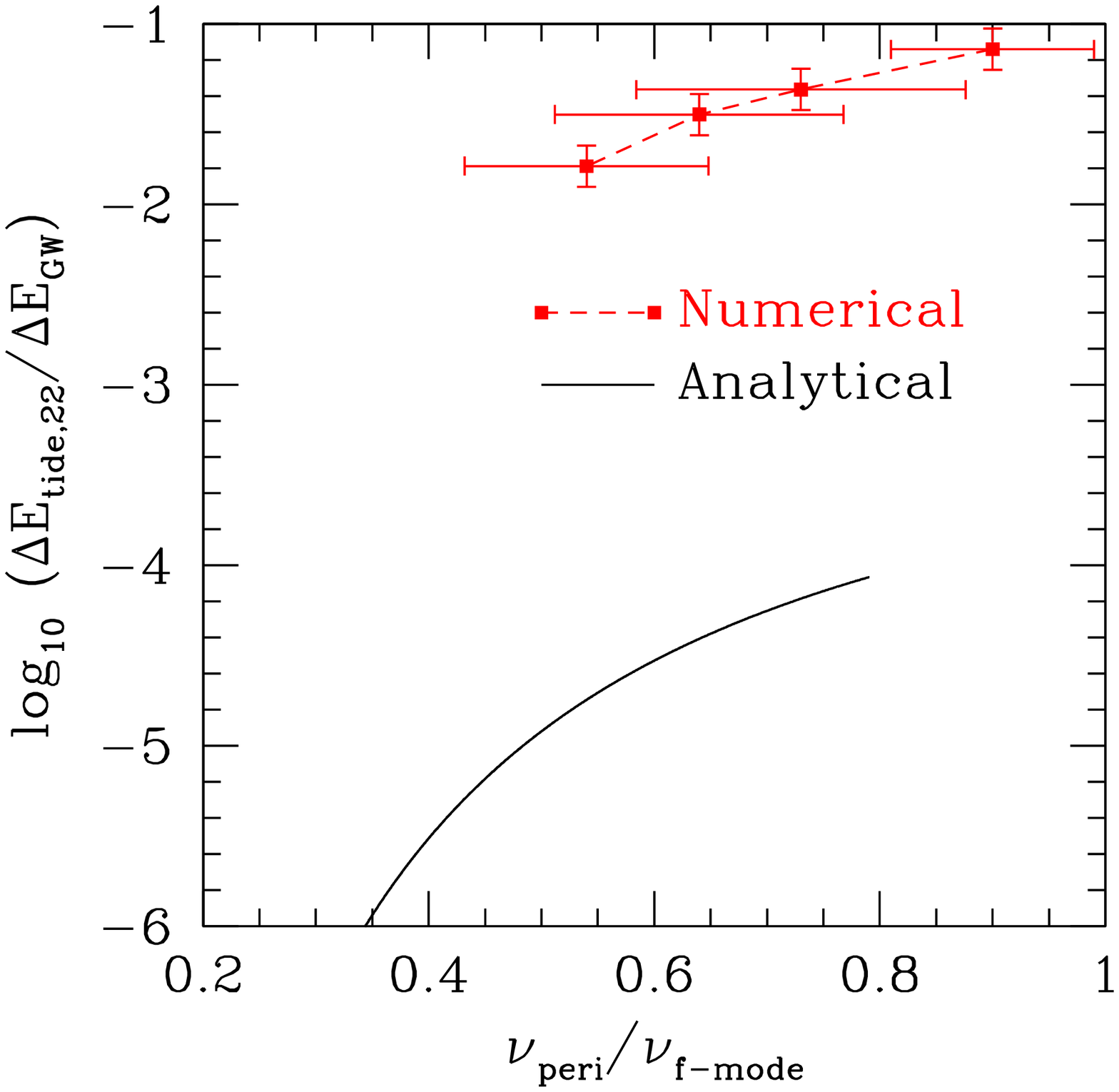}
\vskip-1.0cm
\caption{Comparison of our calculation of equation~(\ref{eq:ratio}) for different ratios of the pericenter frequency $\nu_{\rm peri}$ to the f-mode frequency $\nu_{\rm f-mode}$ (solid black line) with the approximate ratios obtained in two fully general relativistic numerical simulations of \citet{2012PhRvD..86l1501G} (dotted red line). Errorbars in the latter data attempt to roughly quantify systematic uncertainties in wave extraction, isolation of the f-mode contribution and finite resolution via separate numerical tests. In both cases the excitation amplitude decreases with increasing pericenter distance, but the numerical data fall off less steeply than the predictions from linear theory. While some of this discrepancy may well be explained by the challenges inherent to inferring isolated f-mode contributions in this non-linear scenario (see main text for a more detailed discussion), this analysis yields evidence to suggest 
that the fraction of the energy that is deposited in the f-mode is substantially larger than what 
we find in our linear analysis. We suspect that strong pericenter precession, which increases the
time during which the f-mode can be resonantly amplified, is what enhances the energy transfer so
dramatically compared with our Newtonian analytic result.
}
\label{fig:I}
\end{center}
\end{figure}

We need to multiply this by the orbital period $P_{\rm orb}=2\pi a^{3/2}/[G(M_1+M_2)]^{1/2}$ to get the energy lost in one orbit.  When we do so, and when we define $R_{\rm min}=a(1-e)$, set $M_1=M_2$, and take the limit $e\rightarrow 1$, we find
\begin{equation}
\Delta E_{GW}\approx 7\pi {(GM_1)^{7/2}M_1\over{c^5 R_{\rm min}^{7/2}}}\; . \end{equation}
To compare this with $\Delta E_{\rm tide,22}$ it is convenient to rewrite both (and take into account the factor of 2 in the tidal energy deposition):
\begin{equation}
\Delta E_{GW}\approx 7\pi\left(GM_1\over{c^2R_1}\right)^{7/2}\left(R_1\over{R_{\rm min}}\right)^{7/2}M_1c^2
\end{equation}
and
\begin{equation}
\Delta E_{\rm tide,22}={9\over{\pi^2}}\left(GM_1\over{c^2R_1}\right)\left(R_1\over{R_{\rm min}}\right)^3I_{22}^2M_1c^2\; .
\end{equation}
For $R_1c^2/GM_1=6$ (appropriate for $R_1=12$~km and $M_1=1.4~M_\odot$), the ratio is then
\begin{equation}
{\Delta E_{\rm tide,22}\over{\Delta E_{\rm GW}}}\approx 4(R_{\rm min}/R_1)^{1/2}I_{22}^2\; . 
\label{eq:ratio}
\end{equation}
Figure~\ref{fig:I} shows our numerical results for the energy ratio.  We also plot a dotted line indicating the rough amplitude obtained from two simulations of \citet{2012PhRvD..86l1501G}, which shows the results of fully numerical computations of the mode excitation during close encounters.  
The neutron star binary data used in these numerical relativity simulations were obtained using the BAM code of \citet{2008PhRvD..77b4027B} and its hydro extension \citep{2011PhRvD..84d4012T}.  

There are several differences between the work of \citet{2012PhRvD..86l1501G} and ours; for example, they use a polytrope instead of our constant-density star, and pericenter precession is automatically included. We must also note that in the linear theory the orbital and f-mode contributions to the total gravitational wave output can be computed independently, whereas in the non-linear regime such a clean distinction cannot be made as the encounter occurs at smaller and smaller distances. Our  numerical analysis is based on the gravitational wave luminosity, which shows clear sinusoidal variations after the first encounter. 
Concretely, we fit a curve to all local minima in the f-mode oscillations. The f-mode contribution to the radiated energy is then estimated to be the integral over the total gravitational luminosity (including the f-mode oscillations) minus the integral over the fitted curve (which does not include the f-mode). The orbital contribution on the other hand is given by the integral over the pericenter plus the contributions from the fitted curve (again without the f-mode oscillations).

Apart from the mere distinction of the effects due to f-modes over everything else, there are a number of challenges involved in quantifying the total gravitational wave output in a non-linear, numerical relativity simulation of eccentric neutron star binaries. Here we list the most important aspects based on tests we have run: (i) The structure of the stars of $O(10\,\textrm{km})$ as well as their orbital motion of $O(200\,\textrm{km})$ has to be computed at sufficient accuracy in order to generate the metric fluctuations self-consistently. This challenge is overcome by adaptive mesh refinement techniques and resolution studies available from \citet{2012PhRvD..86l1501G} indicate that the highest resolutions used lead to sufficiently small errors due to finite resolution. 
(ii) Gravitational waves are properly defined only at future null infinity and are approximated at large, but finite radius via the usual method of extracting the Newman-Penrose scalar $\Psi_4$ at a coordinate sphere which is centered on the center of mass. The size of this extraction sphere has to be large enough to engulf the entire gravitational wave source. For highly eccentric binaries the spheres have to be chosen larger in order to take the larger orbital separations during apocenter into account. The situation is even more challenging, when each star individually acts as a gravitational wave source that moves off center with respect to the extraction sphere. This poses additional requirements to the extraction sphere choice and leads to the computational challenge: (iii) Even when the stellar structure and orbits are evolved accurately, see (i), there still is a separate challenge in {\it propagating} gravitational waves with sufficient accuracy out to large radius where they are extracted. Hence the computational grid cannot be made coarser (as is usually done) which makes these simulations more expensive than their quasi-circular cousins.

We further note that the models used in \citet{2012PhRvD..86l1501G} were not specifically designed for this study and that in the meantime better setups and improved methods are available. We intent to investigate these issues with the newest available techniques in the near future. We also plan to analyze the contributions from tides in the phase evolution of the binary, which will add complementary information to the amplitude-based estimates made in this work.

Despite all these caveats in mind, we conclude that our comparison between linear and numerical results does suggest that for $R_{\rm min}\approx 2R_1$ the linear analysis underestimates considerably the actual strength of excitation of the f-modes.  We suspect that the main reason for the discrepancy is that relativistic precession near the pericenter allows the tidal forcing to dramatically enhance the f-mode resonance.

\subsection{Mode Energy Dissipation Channels}

If the energy that goes into the f-mode leaks into other channels, that will naturally reduce the amplitude of gravitational waves from the f-mode.  Two leakage mechanisms to consider are coupling of the f-mode to other oscillation modes (which could increase the amplitude of the mode or spread the signal over many frequencies) and viscous dissipation (which could go into heating the star or could release energy via neutrino emission).

The high-amplitude, nonlinear nature of the process we consider makes it difficult to  make definitive statements about mode coupling, but we note that the simulations of \citet{2012PhRvD..86l1501G} show a constant frequency rather than a spectrum, and the analysis of \citet{2007PhRvD..75h4038P} indicates that although there is mode coupling, it could occur at an amplitude that is orders of magnitude below the original f-mode amplitude. This suggests that mode coupling is unlikely to drain much energy from the originally excited f-mode.

Viscous dissipation is also negligible. \citet{1991ApJ...373..213I} calculated the gravitational radiation and viscosity damping times for the f-modes of non-rotating polytropes. They showed that the gravitational radiation time scale for the $l = m = 2$ f-mode is at least 9 orders of magnitude smaller than the viscosity time scale (bulk viscosity, neutron-neutron interactions and electron-electron scattering were considered).

In conclusion, it seems likely that the overwhelming majority of the energy that is tidally deposited in f-modes by close eccentric encounters will be radiated in gravitational waves at the f-mode frequency, rather than being radiated at other frequencies or dissipated in heat or neutrinos.

\section{DISCUSSION AND CONCLUSIONS}
\label{sec:summary}

We present a framework for the discussion of NS-NS and NS-BH eccentric mode excitation and detection via the gravitational waves that would be observed using the {\it Einstein Telescope} or the {\it Cosmic Explorer}.  The vast volume that could be probed with {\it ET} or {\it CE} is enough that even extremely rare events such as those we consider, in which the pericenter distance is of order the radii of the neutron stars, could be seen up to several tens of times per year.  Such close passages will excite f-mode oscillations in the stars, which will then radiate gravitational waves.  The energy in the f-mode gravitational waves will be small compared with that emerging directly from the point-mass orbits, but given that the detection of the two-body gravitational radiation will provide highly accurate times, directions, and masses for the pericenter passages, we anticipate that dedicated analysis will reveal the frequencies and damping times of the modes.

These modes will carry information about the internal structure of the stars that is complementary to, and in some senses cleaner than, the information that will be carried by the post-merger oscillations of hypermassive neutron stars.  In particular, the excitation of f-modes by close eccentric encounters will occur for much lower temperatures, and for much less differential rotation and thus much less amplification of internal magnetic fields, than are expected in hypermassive neutron stars.  Thus there will be fewer variables to consider when analyzing the implications of the oscillations that are detected. The independent measurement of the moment of inertia $I$ and the Love number $\lambda$ in these encounters will therefore provide an observational opportunity to test directly the I-Love-Q relations \citep{2013PhRvD..88b3009Y}.

There is, of course, considerable work to be performed to take this suggestion from a generalized scenario to a more precise tool to probe neutron stars using gravitational wave data.  For example, the high amplitude perturbation we envision is formally not treatable in a linear analysis.  In addition, the Newtonian approximations that exist in the literature are likely to be reasonable guides, but might not be quantitatively accurate; an indication of this is that the \citet{2012PhRvD..86l1501G} simulations find an f-mode amplitude that is considerably greater than would be expected using the standard weak-perturbation analysis.  One interesting possibility is that the substantial pericenter precession that is expected for close neutron star encounters increases the time when the driving frequency is near the f-mode resonance and thus strengthens the tidal coupling.  To analyze this further will require an extension of tidal theory beyond Newtonian orbits; significant progress on this in the context of quasicircular orbits has been made by \citet{2016PhRvL.116r1101H} and \citet{2016arXiv160801907S}.

\acknowledgements

We thank Sebastian Gaebel, Tanja Hinderer, Ilya Mandel, Dave Tsang, Kent Yagi, and Nico Yunes for valuable discussions.  This work was supported in part by S{\~{a}}o Paulo Research Foundation (FAPESP) grant 2015/20433-4, by joint research workshop award 2015/50421-8 from FAPESP and the University of Maryland, College Park, and by Perimeter Institute for Theoretical Physics. Research at Perimeter Institute is supported by the Government of Canada through the Department of Innovation, Science and Economic Development Canada and by the Province of Ontario through the Ministry of Research, Innovation and Science.

\bibliography{ms}

\begin{thebibliography}{}
\expandafter\ifx\csname natexlab\endcsname\relax\def\natexlab#1{#1}\fi

\bibitem[{{Abadie} {et~al.}(2010){Abadie}, {Abbott}, {Abbott}, {Abernathy},
  {Accadia}, {Acernese}, {Adams}, {Adhikari}, {Ajith}, {Allen}, \&
  et~al.}]{2010CQGra..27q3001A}
{Abadie}, J., {Abbott}, B.~P., {Abbott}, R., {et~al.} 2010, Classical and
  Quantum Gravity, 27, 173001

\bibitem[{{Abbott} {et~al.}(2016){Abbott}, {Abbott}, {Abbott}, {Abernathy},
  {Ackley}, {Adams}, {Addesso}, {Adhikari}, {Adya}, {Affeldt}, \&
  et~al.}]{2016arXiv160708697A}
{Abbott}, B.~P., {Abbott}, R., {Abbott}, T.~D., {et~al.} 2016, ArXiv e-prints,
  arXiv:1607.08697

\bibitem[{{Agathos} {et~al.}(2015){Agathos}, {Meidam}, {Del Pozzo}, {Li},
  {Tompitak}, {Veitch}, {Vitale}, \& {Van Den Broeck}}]{2015PhRvD..92b3012A}
{Agathos}, M., {Meidam}, J., {Del Pozzo}, W., {et~al.} 2015, \prd, 92, 023012

\bibitem[{{Antoniadis} {et~al.}(2013){Antoniadis}, {Freire}, {Wex}, {Tauris},
  {Lynch}, {van Kerkwijk}, {Kramer}, {Bassa}, {Dhillon}, {Driebe}, {Hessels},
  {Kaspi}, {Kondratiev}, {Langer}, {Marsh}, {McLaughlin}, {Pennucci}, {Ransom},
  {Stairs}, {van Leeuwen}, {Verbiest}, \& {Whelan}}]{2013Sci...340..448A}
{Antoniadis}, J., {Freire}, P.~C.~C., {Wex}, N., {et~al.} 2013, Science, 340,
  448

\bibitem[{{Baiotti} {et~al.}(2010){Baiotti}, {Damour}, {Giacomazzo}, {Nagar},
  \& {Rezzolla}}]{2010PhRvL.105z1101B}
{Baiotti}, L., {Damour}, T., {Giacomazzo}, B., {Nagar}, A., \& {Rezzolla}, L.
  2010, Physical Review Letters, 105, 261101

\bibitem[{{Bauswein} \& {Janka}(2012)}]{2012PhRvL.108a1101B}
{Bauswein}, A., \& {Janka}, H.-T. 2012, Physical Review Letters, 108, 011101

\bibitem[{{Bauswein} {et~al.}(2012){Bauswein}, {Janka}, {Hebeler}, \&
  {Schwenk}}]{2012PhRvD..86f3001B}
{Bauswein}, A., {Janka}, H.-T., {Hebeler}, K., \& {Schwenk}, A. 2012, \prd, 86,
  063001

\bibitem[{{Bauswein} {et~al.}(2014){Bauswein}, {Stergioulas}, \&
  {Janka}}]{2014PhRvD..90b3002B}
{Bauswein}, A., {Stergioulas}, N., \& {Janka}, H.-T. 2014, \prd, 90, 023002

\bibitem[{{Bauswein} {et~al.}(2016){Bauswein}, {Stergioulas}, \&
  {Janka}}]{2016EPJA...52...56B}
---. 2016, European Physical Journal A, 52, 56

\bibitem[{{Bernuzzi} {et~al.}(2015){Bernuzzi}, {Nagar}, {Dietrich}, \&
  {Damour}}]{2015PhRvL.114p1103B}
{Bernuzzi}, S., {Nagar}, A., {Dietrich}, T., \& {Damour}, T. 2015, Physical
  Review Letters, 114, 161103

\bibitem[{{Bernuzzi} {et~al.}(2012){Bernuzzi}, {Nagar}, {Thierfelder}, \&
  {Br{\"u}gmann}}]{2012PhRvD..86d4030B}
{Bernuzzi}, S., {Nagar}, A., {Thierfelder}, M., \& {Br{\"u}gmann}, B. 2012,
  \prd, 86, 044030

\bibitem[{{Berry} {et~al.}(2015){Berry}, {Mandel}, {Middleton}, {Singer},
  {Urban}, {Vecchio}, {Vitale}, {Cannon}, {Farr}, {Farr}, {Graff}, {Hanna},
  {Haster}, {Mohapatra}, {Pankow}, {Price}, {Sidery}, \&
  {Veitch}}]{2015ApJ...804..114B}
{Berry}, C.~P.~L., {Mandel}, I., {Middleton}, H., {et~al.} 2015, \apj, 804, 114

\bibitem[{{Br{\"u}gmann} {et~al.}(2008){Br{\"u}gmann}, {Gonz{\'a}lez},
  {Hannam}, {Husa}, {Sperhake}, \& {Tichy}}]{2008PhRvD..77b4027B}
{Br{\"u}gmann}, B., {Gonz{\'a}lez}, J.~A., {Hannam}, M., {et~al.} 2008, \prd,
  77, 024027

\bibitem[{Chirenti {et~al.}(2015)Chirenti, de~Souza, \&
  Kastaun}]{Chirenti:2015dda}
Chirenti, C., de~Souza, G.~H., \& Kastaun, W. 2015, Phys. Rev., D91, 044034

\bibitem[{{Del Pozzo} {et~al.}(2013){Del Pozzo}, {Li}, {Agathos}, {Van Den
  Broeck}, \& {Vitale}}]{2013PhRvL.111g1101D}
{Del Pozzo}, W., {Li}, T.~G.~F., {Agathos}, M., {Van Den Broeck}, C., \&
  {Vitale}, S. 2013, Physical Review Letters, 111, 071101

\bibitem[{{East} {et~al.}(2012){East}, {Pretorius}, \&
  {Stephens}}]{2012PhRvD..85l4009E}
{East}, W.~E., {Pretorius}, F., \& {Stephens}, B.~C. 2012, \prd, 85, 124009

\bibitem[{{Favata}(2014)}]{2014PhRvL.112j1101F}
{Favata}, M. 2014, Physical Review Letters, 112, 101101

\bibitem[{{Feroci} {et~al.}(2012){Feroci}, {Stella}, {van der Klis},
  {Courvoisier}, {Hernanz}, {Hudec}, {Santangelo}, {Walton}, {Zdziarski},
  {Barret}, {Belloni}, {Braga}, {Brandt}, {Budtz-J{\o}rgensen}, {Campana}, {den
  Herder}, {Huovelin}, {Israel}, {Pohl}, {Ray}, {Vacchi}, {Zane}, {Argan},
  {Attin{\`a}}, {Bertuccio}, {Bozzo}, {Campana}, {Chakrabarty}, {Costa}, {De
  Rosa}, {Del Monte}, {Di Cosimo}, {Donnarumma}, {Evangelista}, {Haas},
  {Jonker}, {Korpela}, {Labanti}, {Malcovati}, {Mignani}, {Muleri},
  {Rapisarda}, {Rashevsky}, {Rea}, {Rubini}, {Tenzer}, {Wilson-Hodge},
  {Winter}, {Wood}, {Zampa}, {Zampa}, {Abramowicz}, {Alpar}, {Altamirano},
  {Alvarez}, {Amati}, {Amoros}, {Antonelli}, {Artigue}, {Azzarello},
  {Bachetti}, {Baldazzi}, {Barbera}, {Barbieri}, {Basa}, {Baykal}, {Belmont},
  {Boirin}, {Bonvicini}, {Burderi}, {Bursa}, {Cabanac}, {Cackett}, {Caliandro},
  {Casella}, {Chaty}, {Chenevez}, {Coe}, {Collura}, {Corongiu}, {Covino},
  {Cusumano}, {D'Amico}, {Dall'Osso}, {De Martino}, {De Paris}, {Di Persio},
  {Di Salvo}, {Done}, {Dov{\v c}iak}, {Drago}, {Ertan}, {Fabiani}, {Falanga},
  {Fender}, {Ferrando}, {Della Monica Ferreira}, {Fraser}, {Frontera},
  {Fuschino}, {Galvez}, {Gandhi}, {Giommi}, {Godet}, {G{\"o}{\v g}{\"u}{\c s}},
  {Goldwurm}, {G{\"o}tz}, {Grassi}, {Guttridge}, {Hakala}, {Henri}, {Hermsen},
  {Horak}, {Hornstrup}, {in't Zand}, {Isern}, {Kalemci}, {Kanbach}, {Karas},
  {Kataria}, {Kennedy}, {Klochkov}, {Klu{\'z}niak}, {Kokkotas}, {Kreykenbohm},
  {Krolik}, {Kuiper}, {Kuvvetli}, {Kylafis}, {Lattimer}, {Lazzarotto}, {Leahy},
  {Lebrun}, {Lin}, {Lund}, {Maccarone}, {Malzac}, {Marisaldi}, {Martindale},
  {Mastropietro}, {McClintock}, {McHardy}, {Mendez}, {Mereghetti}, {Miller},
  {Mineo}, {Morelli}, {Morsink}, {Motch}, {Motta}, {Mu{\~n}oz-Darias},
  {Naletto}, {Neustroev}, {Nevalainen}, {Olive}, {Orio}, {Orlandini},
  {Orleanski}, {Ozel}, {Pacciani}, {Paltani}, {Papadakis}, {Papitto},
  {Patruno}, {Pellizzoni}, {Petr{\'a}{\v c}ek}, {Petri}, {Petrucci}, {Phlips},
  {Picolli}, {Possenti}, {Psaltis}, {Rambaud}, {Reig}, {Remillard},
  {Rodriguez}, {Romano}, {Romanova}, {Schanz}, {Schmid}, {Segreto}, {Shearer},
  {Smith}, {Smith}, {Soffitta}, {Stergioulas}, {Stolarski}, {Stuchlik},
  {Tiengo}, {Torres}, {T{\"o}r{\"o}k}, {Turolla}, {Uttley}, {Vaughan},
  {Vercellone}, {Waters}, {Watts}, {Wawrzaszek}, {Webb}, {Wilms}, {Zampieri},
  {Zezas}, \& {Ziolkowski}}]{2012ExA....34..415F}
{Feroci}, M., {Stella}, L., {van der Klis}, M., {et~al.} 2012, Experimental
  Astronomy, 34, 415

\bibitem[{{Flanagan} \& {Hinderer}(2008)}]{2008PhRvD..77b1502F}
{Flanagan}, {\'E}.~{\'E}., \& {Hinderer}, T. 2008, \prd, 77, 021502

\bibitem[{{Fonseca} {et~al.}(2016){Fonseca}, {Pennucci}, {Ellis}, {Stairs},
  {Nice}, {Ransom}, {Demorest}, {Arzoumanian}, {Crowter}, {Dolch}, {Ferdman},
  {Gonzalez}, {Jones}, {Jones}, {Lam}, {Levin}, {McLaughlin}, {Stovall},
  {Swiggum}, \& {Zhu}}]{2016arXiv160300545F}
{Fonseca}, E., {Pennucci}, T.~T., {Ellis}, J.~A., {et~al.} 2016, \apj, 832, 167

\bibitem[{{Gair} {et~al.}(2005){Gair}, {Kennefick}, \&
  {Larson}}]{2005PhRvD..72h4009G}
{Gair}, J.~R., {Kennefick}, D.~J., \& {Larson}, S.~L. 2005, \prd, 72, 084009

\bibitem[{{Gendreau} {et~al.}(2012){Gendreau}, {Arzoumanian}, \&
  {Okajima}}]{2012SPIE.8443E..13G}
{Gendreau}, K.~C., {Arzoumanian}, Z., \& {Okajima}, T. 2012, in \procspie, Vol.
  8443, Space Telescopes and Instrumentation 2012: Ultraviolet to Gamma Ray,
  844313

\bibitem[{{Gold} {et~al.}(2012){Gold}, {Bernuzzi}, {Thierfelder},
  {Br{\"u}gmann}, \& {Pretorius}}]{2012PhRvD..86l1501G}
{Gold}, R., {Bernuzzi}, S., {Thierfelder}, M., {Br{\"u}gmann}, B., \&
  {Pretorius}, F. 2012, \prd, 86, 121501

\bibitem[{{G{\"u}ltekin} {et~al.}(2004){G{\"u}ltekin}, {Miller}, \&
  {Hamilton}}]{2004ApJ...616..221G}
{G{\"u}ltekin}, K., {Miller}, M.~C., \& {Hamilton}, D.~P. 2004, \apj, 616, 221

\bibitem[{{G{\"u}ltekin} {et~al.}(2006){G{\"u}ltekin}, {Miller}, \&
  {Hamilton}}]{2006ApJ...640..156G}
---. 2006, \apj, 640, 156

\bibitem[{{Hinderer}(2008)}]{2008ApJ...677.1216H}
{Hinderer}, T. 2008, \apj, 677, 1216

\bibitem[{{Hinderer} {et~al.}(2010){Hinderer}, {Lackey}, {Lang}, \&
  {Read}}]{2010PhRvD..81l3016H}
{Hinderer}, T., {Lackey}, B.~D., {Lang}, R.~N., \& {Read}, J.~S. 2010, \prd,
  81, 123016

\bibitem[{{Hinderer} {et~al.}(2016){Hinderer}, {Taracchini}, {Foucart},
  {Buonanno}, {Steinhoff}, {Duez}, {Kidder}, {Pfeiffer}, {Scheel}, {Szilagyi},
  {Hotokezaka}, {Kyutoku}, {Shibata}, \& {Carpenter}}]{2016PhRvL.116r1101H}
{Hinderer}, T., {Taracchini}, A., {Foucart}, F., {et~al.} 2016, Physical Review
  Letters, 116, 181101

\bibitem[{{Ho} \& {Lai}(1999)}]{1999MNRAS.308..153H}
{Ho}, W.~C.~G., \& {Lai}, D. 1999, \mnras, 308, 153

\bibitem[{{Hotokezaka} {et~al.}(2013){Hotokezaka}, {Kyutoku}, \&
  {Shibata}}]{2013PhRvD..87d4001H}
{Hotokezaka}, K., {Kyutoku}, K., \& {Shibata}, M. 2013, \prd, 87, 044001

\bibitem[{{Huerta} \& {Brown}(2013)}]{2013PhRvD..87l7501H}
{Huerta}, E.~A., \& {Brown}, D.~A. 2013, \prd, 87, 127501

\bibitem[{{Ipser} \& {Lindblom}(1991)}]{1991ApJ...373..213I}
{Ipser}, J.~R., \& {Lindblom}, L. 1991, \apj, 373, 213

\bibitem[{{Kocsis} {et~al.}(2006){Kocsis}, {G{\'a}sp{\'a}r}, \&
  {M{\'a}rka}}]{2006ApJ...648..411K}
{Kocsis}, B., {G{\'a}sp{\'a}r}, M.~E., \& {M{\'a}rka}, S. 2006, \apj, 648, 411

\bibitem[{{Kocsis} \& {Levin}(2012)}]{2012PhRvD..85l3005K}
{Kocsis}, B., \& {Levin}, J. 2012, \prd, 85, 123005

\bibitem[{{Kokkotas} \& {Schafer}(1995)}]{1995MNRAS.275..301K}
{Kokkotas}, K.~D., \& {Schafer}, G. 1995, \mnras, 275, 301

\bibitem[{{Lackey} {et~al.}(2014){Lackey}, {Kyutoku}, {Shibata}, {Brady}, \&
  {Friedman}}]{2014PhRvD..89d3009L}
{Lackey}, B.~D., {Kyutoku}, K., {Shibata}, M., {Brady}, P.~R., \& {Friedman},
  J.~L. 2014, \prd, 89, 043009

\bibitem[{{Lau} {et~al.}(2010){Lau}, {Leung}, \& {Lin}}]{2010ApJ...714.1234L}
{Lau}, H.~K., {Leung}, P.~T., \& {Lin}, L.~M. 2010, \apj, 714, 1234

\bibitem[{{Lindblom}(1992)}]{1992ApJ...398..569L}
{Lindblom}, L. 1992, \apj, 398, 569

\bibitem[{{Miller}(2013)}]{2013arXiv1312.0029M}
{Miller}, M.~C. 2013, ArXiv e-prints, arXiv:1312.0029

\bibitem[{{Miller} \& {Lamb}(2016)}]{2016EPJA...52...63M}
{Miller}, M.~C., \& {Lamb}, F.~K. 2016, European Physical Journal A, 52, 63

\bibitem[{{O'Leary} {et~al.}(2009){O'Leary}, {Kocsis}, \&
  {Loeb}}]{2009MNRAS.395.2127O}
{O'Leary}, R.~M., {Kocsis}, B., \& {Loeb}, A. 2009, \mnras, 395, 2127

\bibitem[{{Pappas}(2015)}]{2015MNRAS.454.4066P}
{Pappas}, G. 2015, \mnras, 454, 4066

\bibitem[{{Passamonti} {et~al.}(2007){Passamonti}, {Stergioulas}, \&
  {Nagar}}]{2007PhRvD..75h4038P}
{Passamonti}, A., {Stergioulas}, N., \& {Nagar}, A. 2007, \prd, 75, 084038

\bibitem[{{Peters}(1964)}]{1964PhRv..136.1224P}
{Peters}, P.~C. 1964, Physical Review, 136, 1224

\bibitem[{{Press} \& {Teukolsky}(1977)}]{1977ApJ...213..183P}
{Press}, W.~H., \& {Teukolsky}, S.~A. 1977, \apj, 213, 183

\bibitem[{{Punturo} {et~al.}(2010){Punturo}, {Abernathy}, {Acernese}, {Allen},
  {Andersson}, {Arun}, {Barone}, {Barr}, {Barsuglia}, {Beker}, {Beveridge},
  {Birindelli}, {Bose}, {Bosi}, {Braccini}, {Bradaschia}, {Bulik}, {Calloni},
  {Cella}, {Chassande Mottin}, {Chelkowski}, {Chincarini}, {Clark}, {Coccia},
  {Colacino}, {Colas}, {Cumming}, {Cunningham}, {Cuoco}, {Danilishin},
  {Danzmann}, {De Luca}, {De Salvo}, {Dent}, {De Rosa}, {Di Fiore}, {Di
  Virgilio}, {Doets}, {Fafone}, {Falferi}, {Flaminio}, {Franc}, {Frasconi},
  {Freise}, {Fulda}, {Gair}, {Gemme}, {Gennai}, {Giazotto}, {Glampedakis},
  {Granata}, {Grote}, {Guidi}, {Hammond}, {Hannam}, {Harms}, {Heinert},
  {Hendry}, {Heng}, {Hennes}, {Hild}, {Hough}, {Husa}, {Huttner}, {Jones},
  {Khalili}, {Kokeyama}, {Kokkotas}, {Krishnan}, {Lorenzini}, {L{\"u}ck},
  {Majorana}, {Mandel}, {Mandic}, {Martin}, {Michel}, {Minenkov}, {Morgado},
  {Mosca}, {Mours}, {M{\"u}ller--Ebhardt}, {Murray}, {Nawrodt}, {Nelson},
  {Oshaughnessy}, {Ott}, {Palomba}, {Paoli}, {Parguez}, {Pasqualetti},
  {Passaquieti}, {Passuello}, {Pinard}, {Poggiani}, {Popolizio}, {Prato},
  {Puppo}, {Rabeling}, {Rapagnani}, {Read}, {Regimbau}, {Rehbein}, {Reid},
  {Rezzolla}, {Ricci}, {Richard}, {Rocchi}, {Rowan}, {R{\"u}diger}, {Sassolas},
  {Sathyaprakash}, {Schnabel}, {Schwarz}, {Seidel}, {Sintes}, {Somiya},
  {Speirits}, {Strain}, {Strigin}, {Sutton}, {Tarabrin}, {Th{\"u}ring}, {van
  den Brand}, {van Leewen}, {van Veggel}, {van den Broeck}, {Vecchio},
  {Veitch}, {Vetrano}, {Vicere}, {Vyatchanin}, {Willke}, {Woan}, {Wolfango}, \&
  {Yamamoto}}]{2010CQGra..27s4002P}
{Punturo}, M., {Abernathy}, M., {Acernese}, F., {et~al.} 2010, Classical and
  Quantum Gravity, 27, 194002

\bibitem[{{Quataert} {et~al.}(1996){Quataert}, {Kumar}, \&
  {Ao}}]{1996ApJ...463..284Q}
{Quataert}, E.~J., {Kumar}, P., \& {Ao}, C.~O. 1996, \apj, 463, 284

\bibitem[{{Quinlan} \& {Shapiro}(1989)}]{1989ApJ...343..725Q}
{Quinlan}, G.~D., \& {Shapiro}, S.~L. 1989, \apj, 343, 725

\bibitem[{{Read} {et~al.}(2009){Read}, {Markakis}, {Shibata}, {Ury{\= u}},
  {Creighton}, \& {Friedman}}]{2009PhRvD..79l4033R}
{Read}, J.~S., {Markakis}, C., {Shibata}, M., {et~al.} 2009, \prd, 79, 124033

\bibitem[{{Read} {et~al.}(2013){Read}, {Baiotti}, {Creighton}, {Friedman},
  {Giacomazzo}, {Kyutoku}, {Markakis}, {Rezzolla}, {Shibata}, \&
  {Taniguchi}}]{2013PhRvD..88d4042R}
{Read}, J.~S., {Baiotti}, L., {Creighton}, J.~D.~E., {et~al.} 2013, \prd, 88,
  044042

\bibitem[{{Reisenegger}(1994)}]{1994ApJ...432..296R}
{Reisenegger}, A. 1994, \apj, 432, 296

\bibitem[{{Samsing} {et~al.}(2014){Samsing}, {MacLeod}, \&
  {Ramirez-Ruiz}}]{2014ApJ...784...71S}
{Samsing}, J., {MacLeod}, M., \& {Ramirez-Ruiz}, E. 2014, \apj, 784, 71

\bibitem[{{Smeyers} \& {van Hoolst}(2010)}]{2010ASSL..371.....S}
{Smeyers}, P., \& {van Hoolst}, T., eds. 2010, Astrophysics and Space Science
  Library, Vol. 371, {Linear Isentropic Oscillations of Stars: Theoretical
  Foundations}

\bibitem[{{Steinhoff} {et~al.}(2016){Steinhoff}, {Hinderer}, {Buonanno}, \&
  {Taracchini}}]{2016arXiv160801907S}
{Steinhoff}, J., {Hinderer}, T., {Buonanno}, A., \& {Taracchini}, A. 2016,
  \prd, 94, 104028

\bibitem[{{Stephens} {et~al.}(2011){Stephens}, {East}, \&
  {Pretorius}}]{2011ApJ...737L...5S}
{Stephens}, B.~C., {East}, W.~E., \& {Pretorius}, F. 2011, \apjl, 737, L5

\bibitem[{{Stone}(2016)}]{2016EPJA...52...66S}
{Stone}, J.~R. 2016, European Physical Journal A, 52, 66

\bibitem[{{Tai} {et~al.}(2014){Tai}, {McWilliams}, \&
  {Pretorius}}]{2014PhRvD..90j3001T}
{Tai}, K.~S., {McWilliams}, S.~T., \& {Pretorius}, F. 2014, \prd, 90, 103001

\bibitem[{{Takami} {et~al.}(2014){Takami}, {Rezzolla}, \&
  {Baiotti}}]{2014PhRvL.113i1104T}
{Takami}, K., {Rezzolla}, L., \& {Baiotti}, L. 2014, Physical Review Letters,
  113, 091104

\bibitem[{{Takami} {et~al.}(2015){Takami}, {Rezzolla}, \&
  {Baiotti}}]{2015PhRvD..91f4001T}
---. 2015, \prd, 91, 064001

\bibitem[{{The LIGO Scientific Collaboration} {et~al.}(2016){The LIGO
  Scientific Collaboration}, {the Virgo Collaboration}, {Abbott}, {Abbott},
  {Abbott}, {Abernathy}, {Acernese}, {Ackley}, {Adams}, {Adams}, \&
  et~al.}]{2016arXiv160707456T}
{The LIGO Scientific Collaboration}, {the Virgo Collaboration}, {Abbott},
  B.~P., {et~al.} 2016, ArXiv e-prints, arXiv:1607.07456

\bibitem[{{Thierfelder} {et~al.}(2011){Thierfelder}, {Bernuzzi}, \&
  {Br{\"u}gmann}}]{2011PhRvD..84d4012T}
{Thierfelder}, M., {Bernuzzi}, S., \& {Br{\"u}gmann}, B. 2011, \prd, 84, 044012

\bibitem[{{Tsang}(2013)}]{2013ApJ...777..103T}
{Tsang}, D. 2013, \apj, 777, 103

\bibitem[{{Tsang} \& {Pappas}(2016)}]{2016ApJ...818L..11T}
{Tsang}, D., \& {Pappas}, G. 2016, \apjl, 818, L11

\bibitem[{Tsui \& Leung(2005)}]{Tsui:2005zf}
Tsui, L.~K., \& Leung, P.-T. 2005, Phys. Rev. Lett., 95, 151101

\bibitem[{{Turner}(1977)}]{1977ApJ...216..914T}
{Turner}, M. 1977, \apj, 216, 914

\bibitem[{{Wade} {et~al.}(2014){Wade}, {Creighton}, {Ochsner}, {Lackey},
  {Farr}, {Littenberg}, \& {Raymond}}]{2014PhRvD..89j3012W}
{Wade}, L., {Creighton}, J.~D.~E., {Ochsner}, E., {et~al.} 2014, \prd, 89,
  103012

\bibitem[{{Watts} {et~al.}(2016){Watts}, {Andersson}, {Chakrabarty}, {Feroci},
  {Hebeler}, {Israel}, {Lamb}, {Miller}, {Morsink}, {{\"O}zel}, {Patruno},
  {Poutanen}, {Psaltis}, {Schwenk}, {Steiner}, {Stella}, {Tolos}, \& {van der
  Klis}}]{2016RvMP...88b1001W}
{Watts}, A.~L., {Andersson}, N., {Chakrabarty}, D., {et~al.} 2016, Reviews of
  Modern Physics, 88, 021001

\bibitem[{{Yagi} \& {Yunes}(2013{\natexlab{a}})}]{2013PhRvD..88b3009Y}
{Yagi}, K., \& {Yunes}, N. 2013{\natexlab{a}}, \prd, 88, 023009

\bibitem[{{Yagi} \& {Yunes}(2013{\natexlab{b}})}]{2013Sci...341..365Y}
---. 2013{\natexlab{b}}, Science, 341, 365

\bibitem[{{Yagi} \& {Yunes}(2014)}]{2014PhRvD..89b1303Y}
---. 2014, \prd, 89, 021303

\bibitem[{{Zhao} {et~al.}(2011){Zhao}, {van den Broeck}, {Baskaran}, \&
  {Li}}]{2011PhRvD..83b3005Z}
{Zhao}, W., {van den Broeck}, C., {Baskaran}, D., \& {Li}, T.~G.~F. 2011, \prd,
  83, 023005

\end{thebibliography}

\end{document}